# Comparison of a modal-decomposition and a grid based field solver for free-electron laser simulations

P. Falgari,[1] H.P. Freund,[2] and D.L.A.G, Grimminck[3]

[1]LIME B.V., Eindhoven, the Netherlands

[2]Department of Electrical and Computer Engineering, University of New Mexico, Albuquerque, USA

[3]ASML B.V., Veldhoven, the Netherlands

**Abstract**

We present a comparison of the modal-decomposition based field solver MINERVA to the grid-based solver GENESIS. The modal decomposition used employs a set of fixed-width, zero-curvature Gaussian modes, as opposed to the more-commonly used basis of vacuum-diffraction eigenmodes. Besides providing the dynamical equations for the field modal amplitudes, we give first-principle estimates of the number of modes required to achieve completeness in terms of field and source modal representation as a function of the electron-current and mode size. As a practical application of this new framework, we present steady-state numerical results for several configurations. The numerical results show a good agreement between the modal expansion and the grid-based solver, not only in terms of the total energy produced, but also in terms of the transverse radiation profile.

# 1. Introduction

In modelling the interaction between electrons and electromagnetic radiation in a free-electron laser (FEL), the latter has been often treated in the so-called slowly-varying envelope approximation (SVEA), in which the radiation vector potential $\delta \mathbf{A}$ (or the corresponding electric and magnetic fields) are approximated by the product of a fast-varying, forward-moving plane wave and an amplitude with a dependence on the transverse coordinates and a *weak* dependence on the longitudinal and time coordinates,

$$\delta \mathbf{A}(r,\theta,z,t) = \delta \hat{\mathbf{A}}(r,\theta,z,t) \exp\left[i(kz - \omega t)\right], \qquad (1)$$

with

$$\frac{\partial}{\partial z} \delta \hat{\mathbf{A}} \ll k \delta \hat{\mathbf{A}}, \qquad \frac{\partial}{\partial t} \delta \hat{\mathbf{A}} \ll \omega \delta \hat{\mathbf{A}}. \qquad (2)$$

Assumption (2), combined with the Maxwell's equations for the vector potential, leads to the so-called *paraxial wave equation*, which reads (in Gaussian units)

$$\left(2i\frac{\omega}{c^2}\frac{\partial}{\partial t} + 2ik\frac{\partial}{\partial z} + \nabla_\perp^2\right)\delta \hat{\mathbf{A}} = -\frac{4\pi}{c}\delta \hat{\mathbf{J}}, \qquad (3)$$

where $\delta \hat{\mathbf{J}}$ is the slowly-varying component of the electron source current, where $\delta \hat{\mathbf{J}} = \delta \mathbf{J} \exp(-ikz + i\omega t)$. In the SVEA framework, the paraxial wave equation is further simplified by removing the fast-varying part of the field and source current. This is achieved by averaging the equation over the fast time/space scale, thereby yielding

$$\left(2i\frac{\omega}{c^2}\frac{\partial}{\partial t} + 2ik\frac{\partial}{\partial z} + \nabla_\perp^2\right)\delta \hat{\mathbf{A}} = -\frac{4\pi}{c}\langle \delta \hat{\mathbf{J}} \rangle,$$

(4)
where

$$\langle \delta \hat{\mathbf{J}}(r,\theta,z,t) \rangle = \frac{1}{\lambda}\int_{z-\lambda/2}^{z+\lambda/2} dz' \delta \hat{\mathbf{J}}(r,\theta,z',t), \qquad (5)$$

with $\lambda = 2\pi/k$. Note that the left-hand side of the equation is not affected by the averaging over one wavelength since $\delta \hat{\mathbf{A}}$ depends weakly on $z$ and can be treated as constant over distances of order $\lambda$.

Two different approaches have been employed in the literature to solve Eq. (4). In the first, which is implemented for example in the field solver GENESIS [1], the transverse dependence of the field is treated by discretizing the field on a two-dimensional grid. The Laplacian operator in Eq. (4) is thus replaced by its discretized version. In the second approach, which is discussed in this work and implemented in the field solver MINERVA [2], the field is decomposed over a complete set of transverse modes, and the Laplacian operator can be computed analytically. Furthermore, the single paraxial equation (4) is replaced by a set of coupled dynamical equations for the field mode amplitudes.

While comparisons of the predictions of the field intensity have been done in the past [3], to our knowledge no rigorous comparison of the transverse intensity profile for grid-based and modal-decomposition-based solvers has been done to date. Such a comparison is presented in this work for a realistic experimental set-up.

## 2. Dynamical Equations for the Optical Field

In ref. [2], the authors introduced a new theoretical framework for the solution of the field dynamical equations based on a decomposition of the vector potential in circularly-polarized Gauss-Laguerre modes

$$\delta \mathbf{A} = \sum_{\substack{n=0 \\ l=-\infty}}^{\infty} e_{nl}(r) \left[ \delta A_{nl}^{(1)} \left( \hat{\mathbf{e}}_x \sin\varphi_l + \hat{\mathbf{e}}_y \cos\varphi_l \right) + \delta A_{nl}^{(2)} \left( \hat{\mathbf{e}}_x \cos\varphi_l - \hat{\mathbf{e}}_y \sin\varphi_l \right) \right], \qquad (6)$$

where

$$\varphi_l = kz + l\theta + \alpha(z)\frac{r^2}{w^2} - \omega t, \qquad (7)$$

is the phase, and

$$e_{nl} = \frac{w_0}{w(z)} \zeta^{|l|} L_n^{|l|}(\zeta^2), \qquad (8)$$

describes the radial variation of the mode amplitude, for which $\zeta = \sqrt{2}r/w$, $L_n^l$ is the associated Laguerre polynomial, and $w$ and $\alpha$ denote the spot size and curvature of the mode for a mode waist of $w_0$. This choice of Gaussian modes is made in order to describe the interaction in a helical undulator which is of the form

$$\mathbf{B}_w(z) = B_w \left( \hat{\mathbf{e}}_x \cos k_w z + \hat{\mathbf{e}}_y \sin k_w z \right), \qquad (9)$$

for an undulator period of $\lambda_w = 2\pi/k_w$ close to the axis of symmetry ($k_w r \ll 1$). In the case of a planar undulator, the Gauss-Hermite mode provide a more natural representation [2]. The mode amplitudes $\delta A_{nl}^{(1,2)}$ and the spot size and curvature are assumed to be slowly-varying function of $z$ and $t$, so that

$$\frac{\partial}{\partial z} \begin{pmatrix} \delta A_{nl}^{(1,2)} \\ w \\ \alpha \end{pmatrix} \ll k \begin{pmatrix} \delta A_{nl}^{(1,2)} \\ w \\ \alpha \end{pmatrix}, \quad \frac{\partial}{\partial t} \begin{pmatrix} \delta A_{nl}^{(1,2)} \\ w \\ \alpha \end{pmatrix} \ll \omega \begin{pmatrix} \delta A_{nl}^{(1,2)} \\ w \\ \alpha \end{pmatrix}. \qquad (10)$$

The dynamical equations for the mode amplitudes are described in ref. [2] subject to the SVEA. To first order in the derivatives, the dynamical equations governing the mode amplitudes are

$$\left( \frac{\partial}{\partial z} + \frac{1}{c}\frac{\partial}{\partial t} \right) \delta A_{nl}^{(1)} + K_{nl} \delta A_{nl}^{(2)} - n \left( X \delta A_{n-1,l}^{(2)} - Y \delta A_{n-1,l}^{(1)} \right)$$
$$- (n+l+1) \left( X \delta A_{n+1,l}^{(2)} + Y \delta A_{n+1,l}^{(1)} \right) = S_{nl}^{(1)}, \qquad (11)$$

$$\left( \frac{\partial}{\partial z} + \frac{1}{c}\frac{\partial}{\partial t} \right) \delta A_{nl}^{(2)} - K_{nl} \delta A_{nl}^{(1)} + n \left( X \delta A_{n-1,l}^{(1)} + Y \delta A_{n-1,l}^{(2)} \right)$$
$$+ (n+l+1) \left( X \delta A_{n+1,l}^{(1)} - Y \delta A_{n+1,l}^{(2)} \right) = S_{nl}^{(2)}, \qquad (12)$$

where the source terms are

$$S_{nl}^{(1)} = -\frac{2}{\pi w_0^2} \frac{n!}{(n+l)!} \int_{t-\pi/\omega}^{t+\pi/\omega} dt' \int_0^{2\pi} d\theta \int_0^{\infty} dr\, re_{nl} \left(\delta J_x \cos\varphi_l - \delta J_y \sin\varphi_l\right), \quad (13)$$

$$S_{nl}^{(2)} = \frac{2}{\pi w_0^2} \frac{n!}{(n+l)!} \int_{t-\pi/\omega}^{t+\pi/\omega} dt' \int_0^{2\pi} d\theta \int_0^{\infty} dr\, re_{nl} \left(\delta J_x \sin\varphi_l + \delta J_y \cos\varphi_l\right). \quad (14)$$

In addition,

$$K_{nl} = (l+2n+1)\left(\frac{\alpha}{w}\frac{dw}{dz} - \frac{1}{2}\frac{d\alpha}{dz} - \frac{1+\alpha^2}{kw^2}\right), \quad (15)$$

$$X = \alpha\frac{dw}{dz} - \frac{1}{2}\frac{d\alpha}{dz} + \frac{1-\alpha^2}{kw^2}, \quad (16)$$

$$Y = \frac{2\alpha}{kw^2} - \frac{dw}{dz}, \quad (17)$$

where we define the convective derivative

$$\frac{d}{dz} = \frac{\partial}{\partial z} + \frac{1}{c}\frac{\partial}{\partial t}. \quad (18)$$

Observe that the $X$ and $Y$ coefficients couple different radial mode numbers, $n$, but that there is no coupling between different azimuthal mode numbers, $l$.

The choice of spot size, $w$, and curvature, $\alpha$, can be arbitrary since the Gauss-Laguerre modes constitute a complete set. In practice, however, numerical simulations require using a limited number subset of the complete modal superposition, so that the results of the simulation will depend upon the particular choice of $w$ and $\alpha$. It is clear that this dependence will decrease as the number of modes in the simulation increases. The choice of the modes that are included forms a balance between two opposing requirements. On the one hand, it is desirable to minimize the number of modes in the simulation in order to keep the computational run time as short as possible. On the other hand, the number of modes should be large enough that inaccuracies arising from the limited number of modes is kept to a minimum. In this regard, it is often found that the number of modes required to achieve an accurate determination of the total saturated power is smaller than that required to obtain an accurate representation of the transverse structure of the optical field. In view of these considerations, an important question is whether it is possible to determine *a priori* the number of modes required to achieve a given accuracy.

One possible choice is the expression of $w$ and $\alpha$ for vacuum-diffraction eigenstates; in particular [4],

$$w(z) = w_0\sqrt{1 + \frac{(z-z_0)^2}{z_R^2}}, \quad \alpha(z) = \frac{z-z_0}{z_R}, \quad (19)$$

where $z_0$ specifies the position of the mode waist, and $z_R = \pi w_0^2/\lambda$ is the Rayleigh range. This choice has the advantage of removing the mode-mode coupling in the dynamical equations because it means that $X = Y = 0$.

As the electrons and the optical field co-propagate through the undulator(s), the optical field is strongly coupled to the electron beam. During the initial exponential growth phase of the interaction, the electron beam acts as an optical fiber to channel the optical field. This is referred to as *optical guiding* in the literature. Once saturation is reached, this guiding ceases and the optical field returns to near-vacuum diffraction. However, because of this guiding, the choice of the vacuum solutions for $w$ and $\alpha$ that describe vacuum diffraction means that a very large number of modes may be required to describe the interaction in both the exponential and post-saturation regimes.

An alternate choice for the basis set is to choose a fixed spot size with zero curvature. This may also require a large number of optical modes, but it has the advantage that vacuum diffraction is not an implicit feature of the modal decomposition. Rather, as we will show in the following discussion, optical guiding and vacuum diffraction will follow from the integration of the dynamical equations. In addition, we will discuss a method to obtain a quantitative estimate of the required number of modes.

## 3. The Fixed Spot Size/Zero Curvature Representation

The fixed spot size/zero curvature representation has been incorporated into the MINERVA simulation code [2]. MINERVA is a three-dimensional, time-dependent free-electron laser simulation code that employs a Gaussian (either Gauss-Hermite or Gauss-Laguerre) modal super-position for the optical field. There are three options available for the field solver:

1. A super-position that employs a basis set including the vacuum solutions for the spot size and curvature.
2. A fixed choice for the spot size and a flat phase front (*i.e.*, $\alpha = 0$).
3. An adaptive eigenmode solver where the spot size and curvature are determined on the fly based upon the strength of the interaction. This is referred to as the Source-Dependent Expansion (SDE) in the literature [5]. The advantage of the SDE is that it minimizes the number of mode required in the super-position; however, the underlying assumption is that the $TEM_{00}$ mode is dominant.

We focus attention in this paper on the second option whereby the basis set has a fixed spot size with zero curvature.

The dynamical equations in this case take the form

$$\left(\frac{\partial}{\partial z}+\frac{1}{c}\frac{\partial}{\partial t}\right)\delta A_{nl}^{(1)} - \frac{l+2n+1}{kw^2}\delta A_{nl}^{(2)} - \frac{n}{kw^2}\delta A_{n-1,l}^{(2)} - \frac{n+l+1}{kw^2}\delta A_{n+1,l}^{(2)} = S_{nl}^{(1)}, \qquad (20)$$

$$\left(\frac{\partial}{\partial z}+\frac{1}{c}\frac{\partial}{\partial t}\right)\delta A_{nl}^{(2)} + \frac{l+2n+1}{kw^2}\delta A_{nl}^{(1)} + \frac{n}{kw^2}\delta A_{n-1,l}^{(1)} + \frac{n+l+1}{kw^2}\delta A_{n+1,l}^{(1)} = S_{nl}^{(2)}. \qquad (21)$$

We will show that this basis set is better suited than the modes employing the vacuum diffraction solutions for $w$ and $\alpha$ to simulate free-electron lasers. It is important to observe that, even though the modes have zero curvature, the completeness of the basis set permits the representation of fields with arbitrary curvature. This is shown in the Appendix for the case of the $TEM_{00}$ mode; however, the proof can be extended to higher-order Gaussian modes.

### 3.1 Completeness of the Source Representation

To quantify the number of required modes needed to achieve a given level of accuracy, it is useful to derive the modal decomposition for the field and current in the simple limiting case of an electron current with a radially-symmetric Gaussian transverse profile and a field represented by a pure TEM$_{00}$ mode. In this simple case the modal decomposition can be computed analytically and the definition of *completeness of the basis* defined in a rigorous way.

The source current is defined as

$$\delta \mathbf{J}(r,\theta,z,t) = -e \sum_{i=1}^{N_e} \mathbf{v}_i(t) \delta[\mathbf{x} - \mathbf{x}_i(t)], \quad (22)$$

where $N_e$ is the number of electrons in the bunch, and $[\mathbf{x}_i(t), \mathbf{v}_i(t)]$ are the three-dimensional position and velocity of the $i^{th}$ electron at time $t$. While the electron current is necessarily granular, and is used in MINERVA simulations, it is often useful to treat it as a continuous distribution. Replacing the velocity by the lowest order contribution due to the motion of the electrons in the undulator, we write the sources as

$$S_{nl}^{(1)} \approx -\frac{2eKc}{\pi\gamma v_\| w^2} \frac{n!}{(l+n)!} \int_{z-\lambda/2}^{z+\lambda/2} dz' \int_0^{2\pi} d\theta \int_0^\infty dr\, r e_{nl}\, n_e(r,\theta,z',t) \cos\left[k(z'-v_\| t)+l\theta\right], \quad (23)$$

$$S_{nl}^{(2)} \approx \frac{2eKc}{\pi\gamma v_\| w^2} \frac{n!}{(l+n)!} \int_{z-\lambda/2}^{z+\lambda/2} dz' \int_0^{2\pi} d\theta \int_0^\infty dr\, r e_{nl}\, n_e(r,\theta,z',t) \sin\left[k(z'-v_\| t)+l\theta\right], \quad (24)$$

where $K = eB_w/m_e c^2 k_w$ is the so-called undulator strength parameter, $v_\|$ is the bulk axial electron velocity, $\gamma$ is the relativistic factor, $e$ and $m_e$ are the electronic charge and rest mass, and $c$ is the speed of light *in vacuo*. The electron density distribution, $n_e$, is normalized such that the integral of the density over all space and time in the bunch is equal to the electron number $N_e$.

We now assume an electron distribution that is azimuthally symmetric and of the form

$$n_e(r,\theta,z,t) = \frac{n_b(z,t)}{2\pi\sigma_b^2} \exp(-r^2/2\sigma_b^2), \quad (25)$$

where $n_b(z,t)$ describes the shape of the electron bunch. It is immediately clear that the sources [Eqs. (23) and (24)] vanish for $l \neq 0$ due to the symmetry of the electron bunch; hence,

$$S_{nl}^{(1)} \approx -\delta_{l,0} \frac{eKc}{2\pi\sigma_b^2 \gamma v_\|} \operatorname{Re}[B(z,t)] I_n, \quad (26)$$

$$S_{nl}^{(2)} \approx -\delta_{l,0} \frac{eKc}{2\pi\sigma_b^2 \gamma v_\|} \operatorname{Im}[B(z,t)] I_n, \quad (27)$$

where

$$B(z,t) = \int_{z-\lambda/2}^{z+\lambda/2} dz'\, n_b(z',t) \exp[-ik(z'-v_\| t)], \quad (28)$$

$$I_n = \frac{4}{w^2} \int_0^\infty dr\, r\, e_{n0}(r) \exp(-r^2/2\sigma_b^2) = \left(\frac{1}{2} + \frac{w^2}{4\sigma_b^2}\right)^{-1} \left[1 - \left(\frac{1}{2} + \frac{w^2}{4\sigma_b^2}\right)^{-1}\right]^n. \quad (29)$$

It can be shown that $\Sigma_{n=0}^{\infty} I_n = 1$, so that the source terms in the limit of an azimuthally symmetric electron bunch reduce to a product of a mode-independent bunching factor $B(z,t)$ and a mode weight $I_n$.

The bunching factor measures the strength of the coupling between the electrons and the optical field and is expected to be vanishingly small near the entrance to the undulator and to grow as the interaction proceeds through the undulator and micro-bunching of the electron beam increases. The $I_n$ coefficient measures the relative strength of the coupling of the electron bunch to a particular optical mode. As a result, the sum of these coefficients for a modal super-position of $N$ modes

$$S_N = \sum_{n=1}^{N} I_n, \qquad (30)$$

can be interpreted as the degree of completeness of the basis set, and where $1 - S_N$ provides a quantitative measure of the contribution of the missing modes.

**3.2 Completeness of the Field Representation**

We now consider the modal decomposition of a circularly-polarized TEM$_{00}$ mode *in vacuo*, which can be written in the form [4]

$$\delta \mathbf{A}(r,\theta,z) = \delta A_0 \frac{\hat{e}_x + i\hat{e}_y}{\sqrt{2}} \frac{w_0}{w} \exp\left(-\frac{r^2}{w^2}\right) \exp\left[i\left(kz + \alpha \frac{r^2}{w^2} - \psi(z)\right)\right] + c.c., \qquad (31)$$

for the vector potential, where the spot size and curvature are given by Eqs. (19), and

$$\psi(z) = \tan^{-1}\left(\frac{z}{z_R}\right). \qquad (32)$$

is the Gouy phase shift. We observe that

$$\int_0^{2\pi} d\theta \int_0^{\infty} dr\, r\, |\delta A(r,\theta,z)|^2 = \frac{\pi}{2} w_0^2 |\delta A_0|^2. \qquad (33)$$

As demonstrated in the Appendix, the modal weight for the constant spot size/zero curvature representation can be expressed as

$$W_n = \int_0^{2\pi} d\theta \int_0^{\infty} dr\, r\, |\delta A_{nl} u_{nl}(r,\theta,z)|^2 = \delta_{l,0} \frac{\pi w_0^2 |\delta A_0|^2}{2} \frac{\bar{w}^2}{w^2} \frac{1}{\chi \chi^*} \left[\left(1 - \frac{1}{\chi}\right)\left(1 - \frac{1}{\chi^*}\right)\right]^n. \qquad (A7)$$

As shown in the Appendix, a summation of these modal weights over all $n$ reproduces the integrated intensity of the TEM$_{00}$ Gaussian mode, *i.e.*,

$$\sum_{n=0}^{\infty} W_n = \frac{\pi}{2} w_0^2 |\delta A_0|^2. \qquad (34)$$

We can define the *degree of completeness* of a modal decomposition $n = 1, \ldots, N$ as

$$S_C = \frac{2}{\pi w_0^2 |\delta A_0|^2} \sum_{n=0}^{N} W_n, \qquad (35)$$

where $1 - S_C(N)$ represents the contribution to the overall intensity due to the "missing" modes. It does not provide information on the point wise convergence of the modal decomposition to the exact result; however, it is a useful measure of a quantitative estimate of the relative contribution of the excluded modes.

### 3.1 Examples

We now present two examples of how one can use (30) and (35) to determine the number of modes required to achieve a certain level of completeness in terms of field and source modal representations. The criterion we use is to require that both $S_N$ and $S_C$ are larger than a certain threshold $S_0$,

$$S_N, S_C > S_0 . \tag{36}$$

For the two examples discussed we require a level of completeness of 99%, *i.e.* $S_0 = 0.99$. Note that $I_n$ and $W_n$ depend on the position $z - z_0$, so we require that (36) is satisfied at each position along the undulator. Beside $z - z_0$, $S_N$ and $S_C$ also depend on the electron beam transverse size, $\sigma_b$, the TEM$_{00}$ mode waist, $w_0$, the radiation wavelength $\lambda$ and the mode size $\bar{w}$. To reduce the number of free parameters in our analysis we assume that the transverse size of the optical field at the waist is equal to the electron beam size so that $w_0 = \sqrt{2}\sigma_b$. This choice is motivated by the fact that the optical field is guided by the electron beam interaction in the exponential gain regime.

The first example under consideration corresponds to an experimental configuration that includes an undulator line that is 40 m in length and generates output radiation at a wavelength of 13.5 nm. The upper plot in Fig. 1 shows the number of modes required to achieve 99% completeness as a function of electron beam size $\sigma_b$ (horizontal axis) and the relative optical mode size $\bar{w}/\sigma_b$ (vertical axis). The lower plots show the optimal choice of $\bar{w}$ for a given value of $\sigma_b$ (where the optimal value is defined as the one which gives 99% completeness with the smallest number of modes) and the corresponding number of modes. It is evident that both the optimal value of $\bar{w}$ and $n$ decrease with $\sigma_b$. This can be understood by consideration that the Rayleigh range of the optical mode is $z_R = \pi w_0^2 / \lambda \approx 2\pi\sigma_b^2 / \lambda$, which implies that diffraction increases with decreasing values of $\sigma_b$. As a consequence, more modes are required to achieve a given level of completeness everywhere. For example, $z_R = 1.68$ m for $\sigma_b = 60$ µm and the optical field undergoes a 24-fold increase in size as it propagates through the undulator. In contrast, as $\sigma_b$ increases, the size difference between the optical field at the waist and at the exit from the undulator decreases, and fewer modes are required to achieve completeness. Furthermore, the optimal size of the mode approaches the size of the optical mode at the waist, $w_0 = \sqrt{2}\sigma_b$. In the limit in which the Rayleigh range is much larger than the length of the undulator, both the optical mode and the electron beam have a constant size along the undulator and a single mode with $\bar{w} = \sqrt{2}\sigma_b$ is enough to achieve completeness.

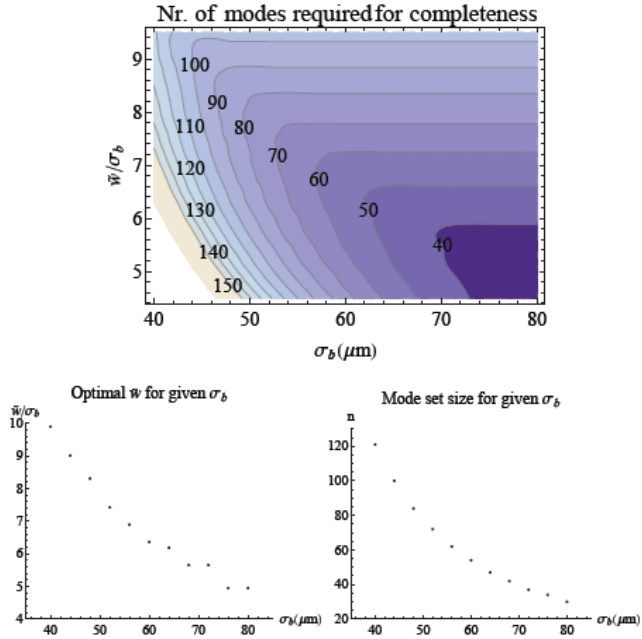

Fig. 1 Modal basis completeness for the 13.5nm wavelength example described in the text. The upper contour plot shows the number of modes required to achieve completeness as a function of electron beam rms size $\sigma_b$ and the relative mode size $\bar{w}/\sigma_b$. The lower plots show the optimal choice of $\bar{w}/\sigma_b$ as a function of $\sigma_b$ (left) and the mode set size for that particular choice of $\bar{w}/\sigma_b$ (right).

The second example under consideration is the Linac Coherent Light Source (LCLS) at the Stanford Linear Accelerator Center [6] in which the undulator line exceeds 100 m in length and the radiation wavelength is 1.5 Å. Plots analogous to Fig. 1 for the configuration under consideration are presented in Fig. 2 to describe the LCLS. The functional dependence of the optimal $\bar{w}$ and the number of required modes is similar to that found for previously, but the number of modes required to achieve completeness for the LCLS is significantly fewer. This is a consequence of the typical Rayleigh range for the value of $\sigma_b$ for the LCLS. For example, $\sigma_b = 20$ mm for the LCLS so that $z_R = 16.8$ m, and the optical mode undergoes a 6-fold expansion along the 100 m of the LCLS undulator, which is four times smaller than the expansion of the optical mode in previous example.

The completeness study presented in this section assumes that the radiation diffracts freely along the whole length of the undulator. In a free-electron laser, where the radiation couples strongly to the electrons and is guided by them, the diffraction is less, so that the results shown in Figs. 1 and 2 should be considered conservative, and the actual number of required modes likely smaller.

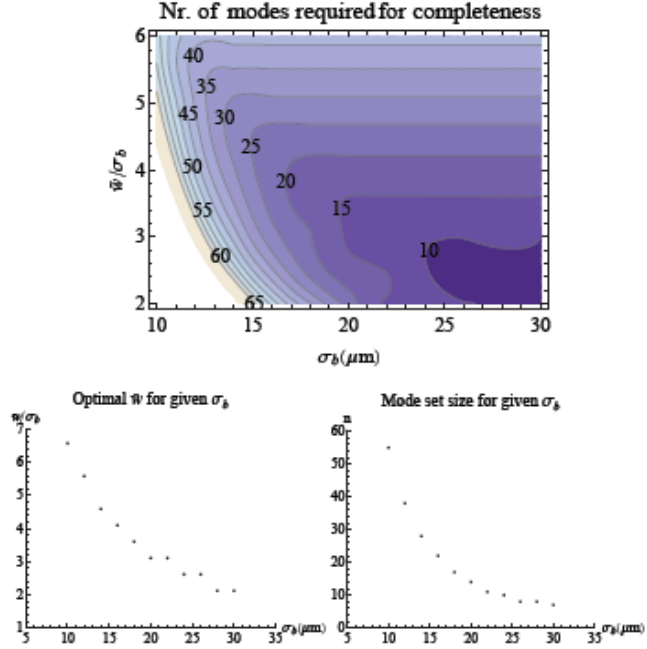

Fig. 2 Modal basis completeness for the LCLS experimental setup. The upper contour plot shows the number of modes required to achieve completeness as a function of electron beam rms size $\sigma_b$ and the relative mode size $\bar{w}/\sigma_b$. The lower plots show the optimal choice of $\bar{w}/\sigma_b$ as a function of $\sigma_b$ (left) and the mode set size for that particular choice of $\bar{w}/\sigma_b$ (right).

| **Electron Beam** | |
|---|---|
| Energy | 1.0 GeV |
| Current | 2500 A |
| Normalized Emittance | 1.4 mm-mrad |
| rms Energy Spread | 0.05% |
| Initial $\sigma_x$, $\sigma_y$ | 87.5 μm, 35.0 μm |
| **Seed Laser** | |
| Wavelength | 13.5 nm |
| Power (TEM$_{00}$) | 40 W |
| Mode Waist | 85 μm |
| Waist Position | 0.4082 (after undulator entrance) |
| **Undulator & FODO Lattice** | |
| Undulator Type | Helical |
| Period | 3.14 cm |
| Field Strength | 5.15 kG |
| Length | 80 periods (1 period transitions) |
| Number of Segments | 12 |
| Quadrupole Gradient | 2.2773 kG/cm |
| Quadrupole Length | 3.14 cm (initial half quad)/6.28 cm |
| Gap Length | 0.8164 m |

Table 1 Summary of the parameters used in the numerical simulations.

In addition, the model used to quantify the number of modes required to achieve a certain level of completeness assumes a cylindrically symmetric current density and field. For a realistic short wavelength (extreme ultraviolet through x-ray) free-electron laser, this assumption is in most cases not justified. The electron beam is typically focused through the undulator by a series of strong focusing quadrupoles, which induces periodic oscillations of the electron envelope along two perpendicular directions. These oscillations result in higher-order azimuthal modes in the modal decomposition of the current, and excite higher-order modes in the field through the source terms in Eqs. (13) and (14). For this reason, it is necessary to include higher order azimuthal modes in the modal decomposition, as discussed in the following section. However, it should be noted that, for typical scenarios, the $l = 0$ modes still represent the dominant contribution to the field modal decomposition, and the completeness study presented in this section is expected to still give a good estimate of the required size of the modal basis.

## 4. Numerical Simulation

We now present a numerical comparison of simulations obtained with MINERVA and GENESIS. For the Minerva simulations we use the fixed-width mode implementation discussed in the previous section. Since we are mainly interested in the comparison of the modal-based and grid-based solvers, we limit ourselves to steady-state simulations, rather than time-dependent ones.

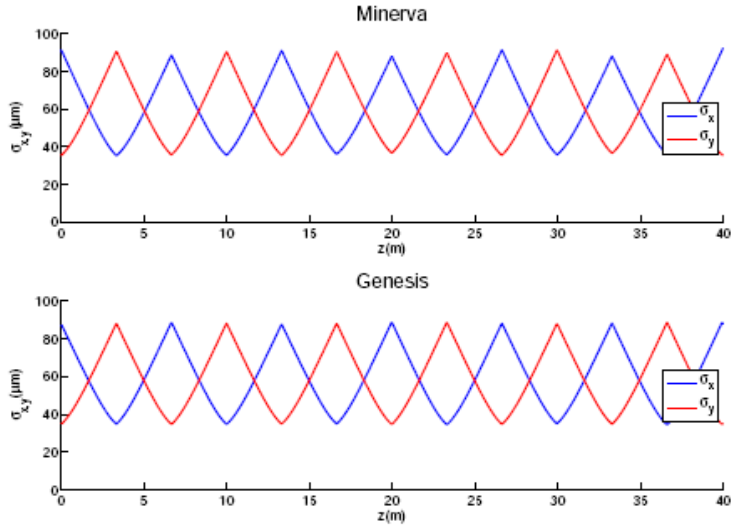

Fig. 3 Results of the matching of the electron beam into the undulator and FODO lattice in MINERVA (top) and GENESIS (bottom). The plots show the rms electron beam sizes in the $x$(blue)- and $y$ (red)- directions as function of position along the undulator.

The parameters used in the simulation are summarized in Table 1. The electron beam energy and current are 1.0 GeV and 2500 A respectively. The normalized emittance is 1.4 mm-mrad in both the $x$- and $y$-directions, and the rms energy spread is 0.05%. The seed laser produces 40 W at a wavelength of 13.5 nm and is focused to a waist size of 85μm at a distance of 0.4082 m after the start if the first undulator. The undulator line consists of 12 helical undulator segments each of which has a period of 3.14 cm, a field strength of 5.15 kG, and is 80 periods in length with one period entry and exit transitions. The separation distance between the undulators is 0.8164 m. The FODO lattice [7] is composed of alternating quadrupoles centered in the gaps between the undulators each of which has a length of 6.28 cm and a field gradient of 2.2773 kG/cm. The electron beam is matched into the undulator/FODO lattice by means of a half-quadrupole located prior to the first undulator and where the electron beam is focused to a waist at that point of 87.5 μm in the $x$-direction and 35.0 μm in the $y$-direction. The transition sections are simulated in MINERVA using the following field profile for the entry transition

$$B_w(z) = B_{w0} \sin^2\left(\frac{\pi z}{2\lambda_w}\right), \tag{37}$$

with a symmetric decrease in the exit transition. The transitions are not simulated in GENESIS, but are treated as additional drift spaces.

The evolution of the beam envelopes in the *x*- and *y*-directions as determined in MINERVA and GENESIS are illustrated in Fig. 3. As shown in the figure, the two codes produce substantially the same results for beam propagation, and show that the beam is well-matched into the FODO lattice.

The aforementioned completeness discussion suggests that 60 modes with $\bar{w}$ = 380 μm are sufficient to achieve a high level of completeness for the case at hand. Therefore, we ran simulations with MINERVA using 20, 40, and 60 purely radial modes in order to study the development of completeness as a function of increasing number of modes. The 6-dimensional electron phase space is modelled in MINERVA with 64800 macro-electrons. The GENESIS simulations used 131072 macro-electrons and a square grid of 5 mm × 5 mm with $901^2$ grid points. A laser seed power of 40 W was injected together with the electrons and focused to a waist of 85 μm at a distance of 0.4082 m into the first undulator.

### 4.1 Vacuum Diffraction

Before presenting the simulation results including the electron beam, we compare the predictions of MINERVA and GENESIS for pure vacuum diffraction. To this end, we simulated the propagation of the pure $TEM_{00}$ laser seed through the undulator. The interaction with the electrons was switched off by setting the undulator field to zero.

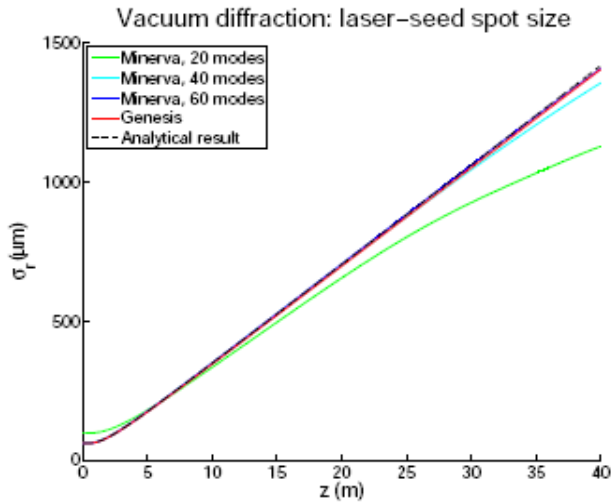

Fig. 4 The rms size of the optical field as a function of *z*, as predicted by GENESIS (red) and MINERVA with 20 green), 40 (cyan), and 60 (blue) modes, and by the exact analytical solution (dashed black).

Figure 4 shows the predicted propagation by MINERVA and GENESIS for the injection of the specified TEM$_{00}$ Gaussian mode and includes the analytic prediction [Eq. (19)] for comparison. It is clear that the convergence of the MINERVA results improves with increasing mode-set size. While the MINERVA result obtained with 20 modes shows a discrepancy from the analytic result as large as 20%, the result obtained with 40 modes differs by at most 5% from the analytic result, and the result obtained with 60 modes differs by less than 1% from the analytic result over the whole undulator length. The GENESIS result has an accuracy similar to that of the 60-mode MINERVA result.

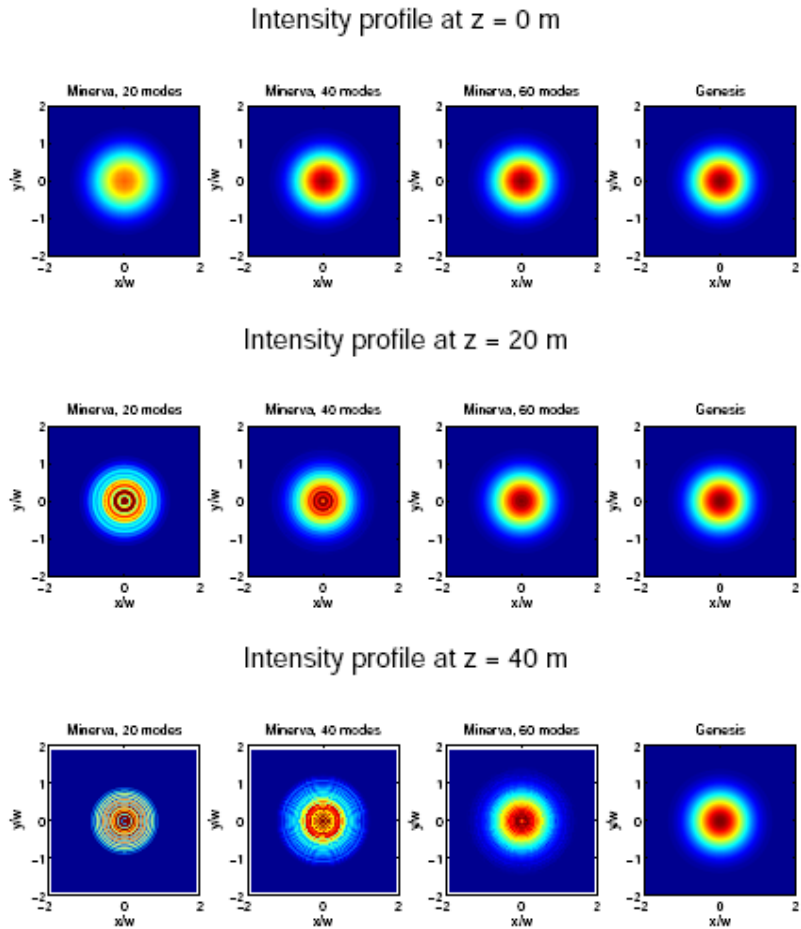

Fig. 5 Intensity profile in the transverse-y plane at z = 0 m, 20 m, and 40 m inside the undulator as predicted by GENESIS and MINERVA (with 20, 40, and 60 modes). The transverse coordinates are normalized to the vacuum diffracting spot size [Eq. (19)].

Figures 5 and 6 show the radiation intensity profile in the transverse plane, and a section along the *x*-axis, at $z = 0$, 20, 40 meters along the undulator. For the intensity profile at $z = 0$, the convergence of the MINERVA results to the analytical result is similar to what seen for the laser spot size. However, at $z = 20$ m and $z = 40$ m, one can clearly see that the point-wise convergence with the number of modes is definitely slower. In particular, the MINERVA results show residual oscillations around the exact analytical predictions. These oscillations arise from the phase difference accumulated by different modes as they diffract through the undulator, which leads to interference patterns. The amplitude of the oscillations decrease visibly with the number of modes, though for 60 modes the difference compared to the exact analytical result is larger than the 1% effect expected from the completeness study presented. It was already pointed out in Section 3.1 that a certain level of predicted accuracy for *integrated* quantities, does not imply the same accuracy for the non-integrated quantities, such as the intensity distribution, so the difference seen between the

MINERVA result with 60 modes and the exact result is not unexpected. Note, however, that for the strongly-coupled scenario which we discuss in Section 4.2, we expect a better accuracy, since the radiation guiding leads to significantly less diffraction along the undulator.

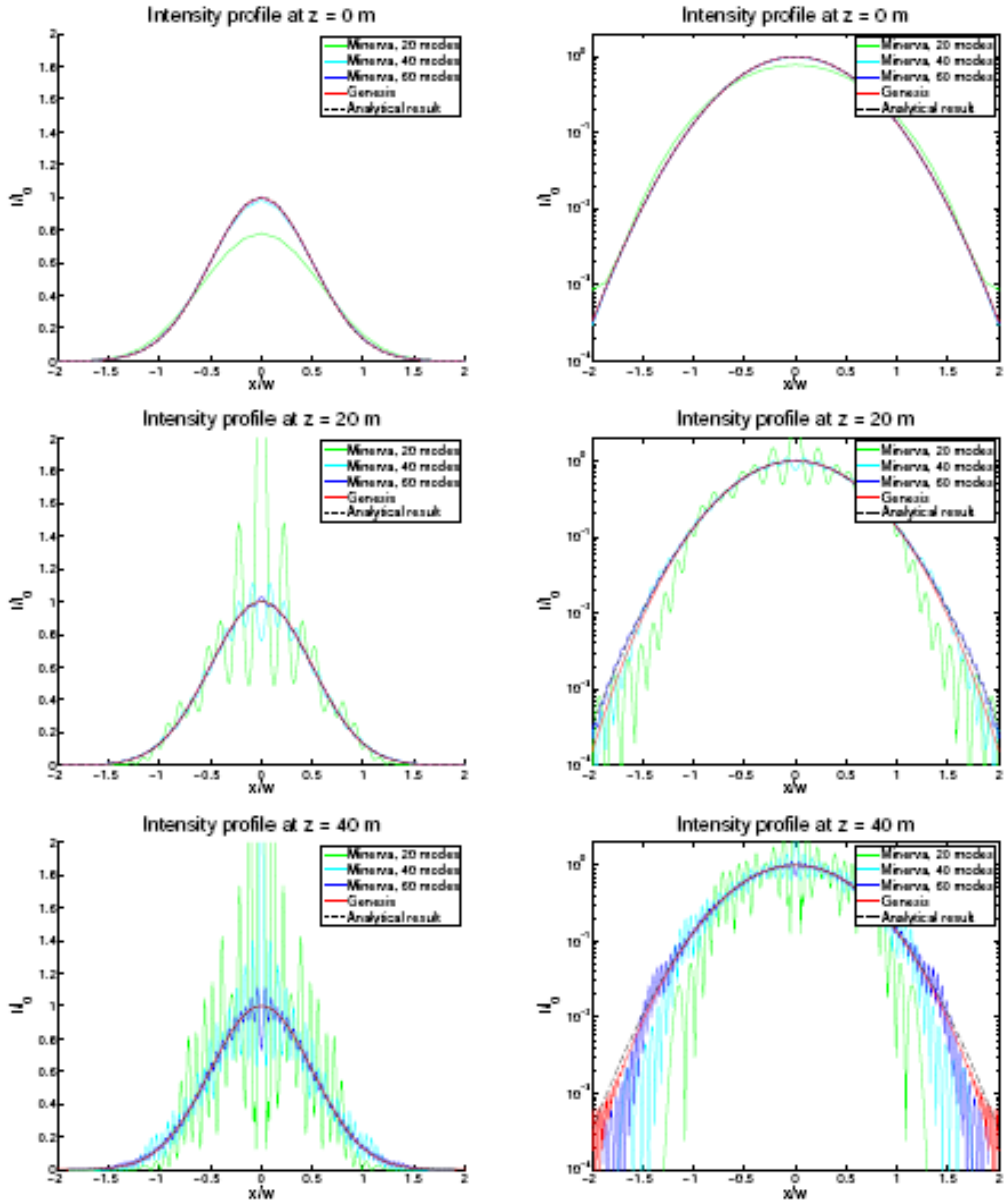

Fig. 6 Intensity profiles along the *x*-axis at $z = 0$, 20, and 40 m inside the undulator as predicted by MINERVA (with 20, 40, and 60 modes), GENESIS, and the analytical result. The transverse coordinates are normalized by the analytic spot size, and the intensity is normalized to the peak value of the analytic result. Plots on the left use a linear vertical axis while those on the right use a logarithmic vertical axis.

To summarize, the constant spot size/zero curvature modal decomposition shows a clear convergence to the exact result for the optical field propagation *in vacuo*. For integrated quantities, like the field spot size, the convergence with the number of modes is as expected from the study presented in Section 3. Pointwise convergence is slower, though, in this respect, the in vacuo case is a worst-case-scenario, and the convergence behavior for the interacting case is expected to be better.

**4.2 Electron Beam Simulations**

We now turn to a comparison of the MINERVA and GENESIS simulations of the electron beam interaction corresponding to the parameters shown in Table 1. As explained above, the MINERVA simulations for the vacuum propagation used 20, 40, and 60 radial modes ($l = 0$, $n = 0$, ... ,19/39/59) with $\bar{w} = 380$ μm. However, in order to capture the azimuthal dependence that may arise from the oscillations in the beam envelope due to the FODO lattice, we added 80 azimuthal modes with $l = \pm1, \pm2$ and $n = 0, ... , 19$.

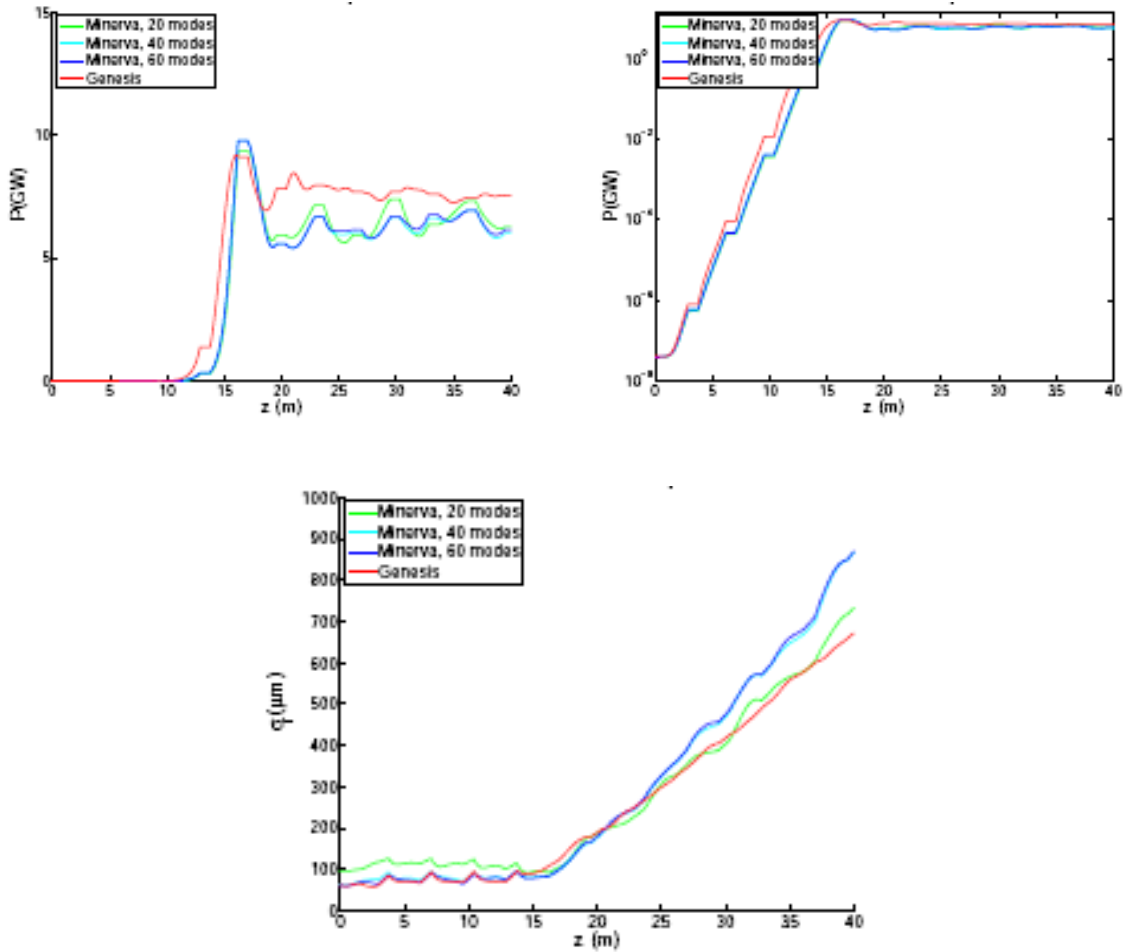

Fig. 7 Total optical power as a function of distance on a linear (upper left) and logarithmic (upper right) scale, and the radiation spot size (lower plot). The plots show predictions by GENESIS (red) and MINERVA with 20 (green), 40 (cyan), and 60 (blue) radial modes. All MINERVA simulations include 80 additional modes with $l = \pm1, \pm2$ and $n = 0, ...,19$.

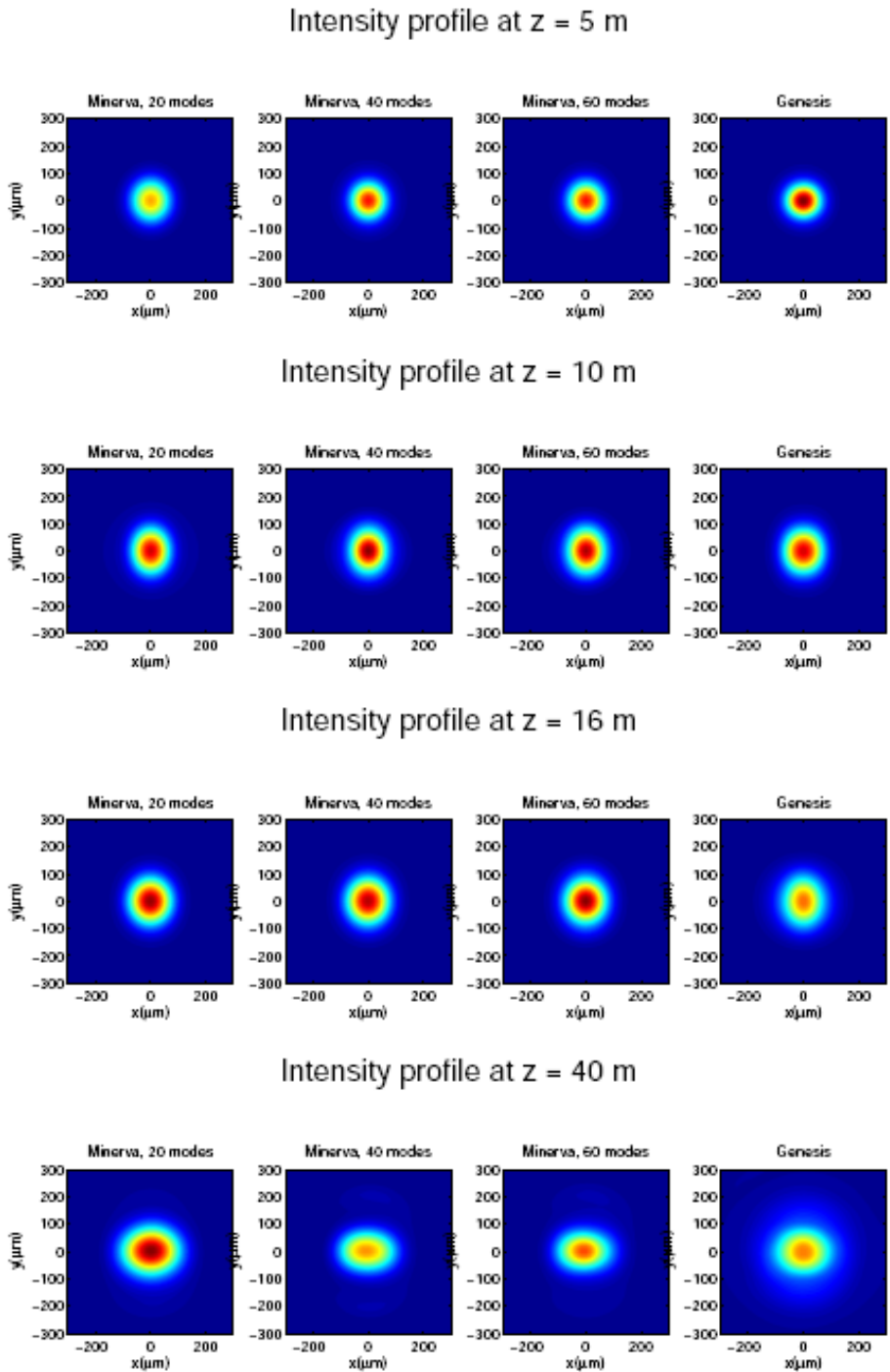

Fig. 8 Normalized intensity profile in the transverse $x - y$ plane predicted by GENESIS and MINERVA with 20, 40 and 60 radial modes at $z =$ 5, 10, 16, 40 m. All MINERVA simulations include 40 additional azimuthal modes with $l =$ 1, 2 and $n =$ 0, ... , 19. Note that the color map is the same for all sub-plots.

Figure 7 shows the total power (upper plots) and the spot size as found using MINERVA and GENESIS. The MINERVA results show a clear convergence with respect to the number of azimuthally symmetric modes since the results using 40 and 60 modes are extremely close over then length of the undulator. Compared to GENESIS, MINERVA predicts a somewhat lower growth rate in the exponential growth regime. However, the two codes predict similar saturation lengths of about 16 m and saturation powers of about 10 GW, which is in good agreement with the predictions based on a parameterization of the free-electron laser interaction given by

Ming Xie [8] after allowance is made for the drift sections between the undulators. After saturation, the two codes show a larger discrepancy, with MINERVA showing a larger reabsorption of power and larger synchrotron-betatron oscillations than GENESIS. Both MINERVA and GENESIS show strong optical guiding in the exponential regime followed by resumed diffraction after saturation. where MINERVA predicts faster diffraction than GENESIS.

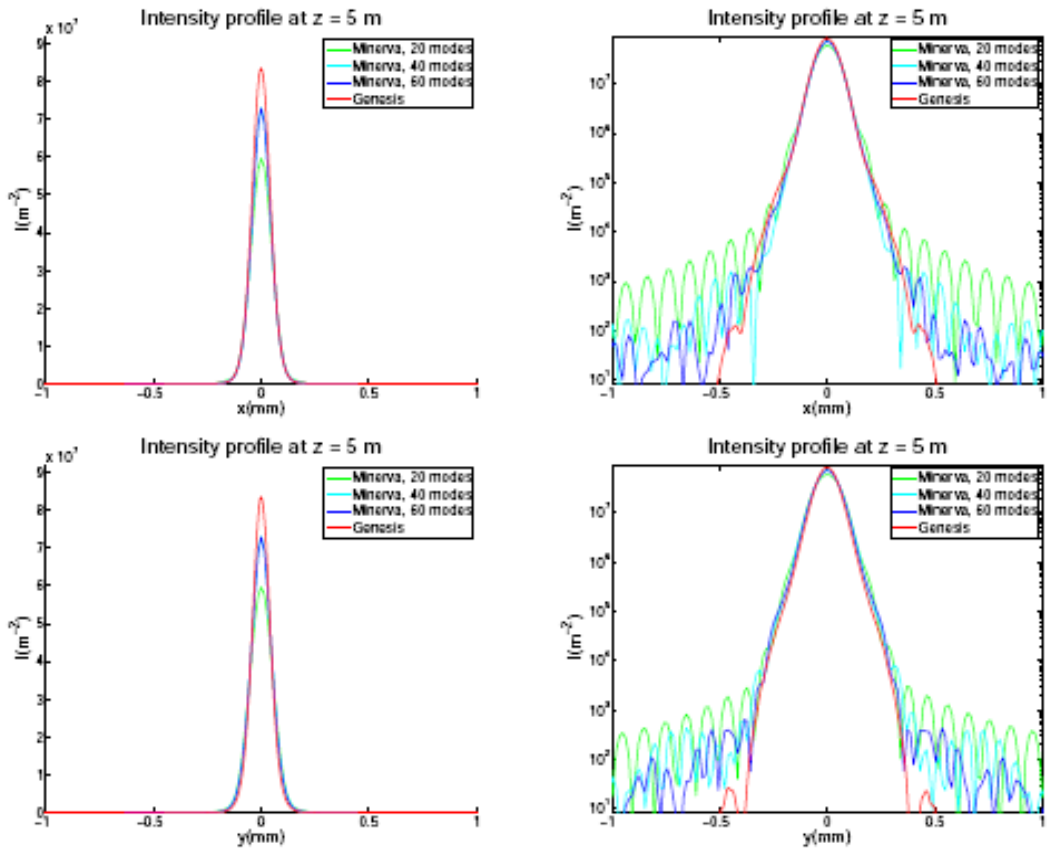

Fig. 9 Sections of the intensity profile at $z$ = 5 m as obtained using GENESIS and MINERVA on a linear (left) and logarithmic (right) scale. The upper plots show the section along the $x$-axis, the lower plots along the $y$-axis.

Figure 8 shows predictions for the intensity profile (normalized to unity) in the transverse plane at 5, 10, 16 and 40 meters after the undulator entrance. Figures 9, 10, 11, 12 show sections of the intensity profile along the $x$- and $z$- axis on a linear (left plots) and logarithmic scale (right plots). The size and shape predicted by MINERVA and GENESIS for the central peak is generally in very good agreement during the exponential regime. Both codes clearly predict an ellipticity of the intensity profile, which can be shown to be correlated to the periodic oscillations of the electron beam induced by the FODO lattice. At $z = 40$ m, which corresponds to a deeply-saturated state, the two codes show much larger differences, with GENESIS notably predicting "shoulders" in the intensity profile, which are absent in the MINERVA result.

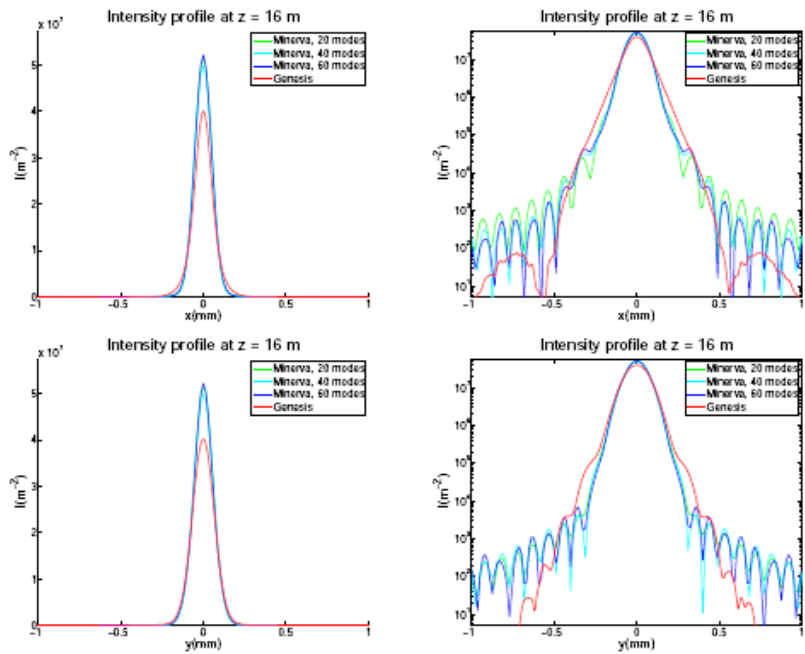

Fig. 10 Sections of the intensity profile at $z = 10$ m as obtained using GENESIS and MINERVA on a linear (left) and logarithmic (right) scale. The upper plots show the section along the *x*-axis, the lower plots along the *y*-axis.

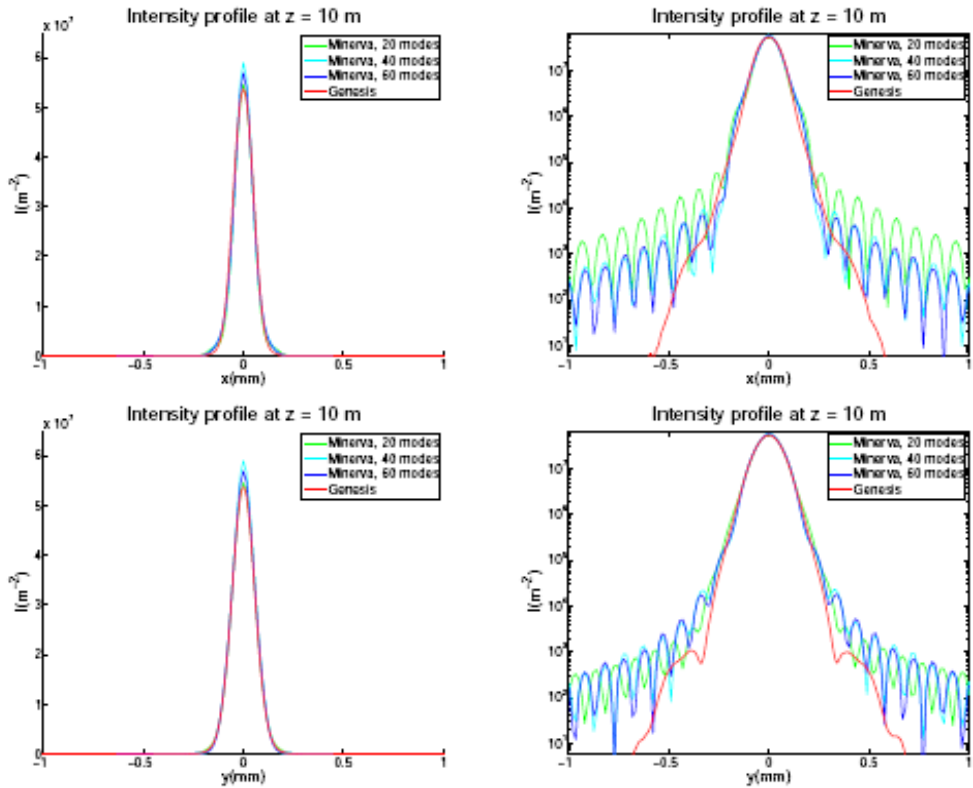

Fig. 11 Sections of the intensity profile at $z = 16$ m as obtained using GENESIS and MINERVA on a linear (left) and logarithmic (right) scale. The upper plots show the section along the *x*-axis, the lower plots along the *y*-axis.

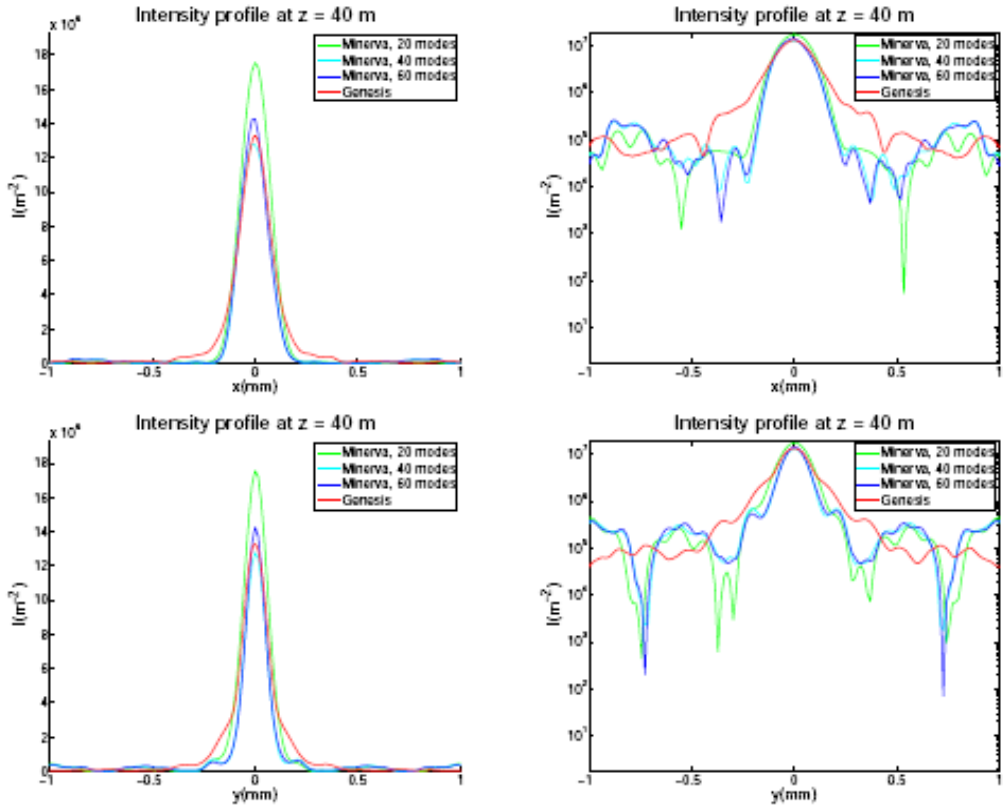

Fig. 12 Sections of the intensity profile at *z* = 40 m as obtained using GENESIS and MINERVA on a linear (left) and logarithmic (right) scale. The upper plots show the section along the *x*-axis, the lower plots along the *y*-axis.

## 5. Summary and Discussion

In this paper, we have presented an analysis of the utility of using a modal expansion based on a fixed spot size/zero curvature representation of Gaussian optical modes in the simulation of short wavelength free-electron lasers. Starting with a general description of the dynamical equations of the optical field for an arbitrary Gaussian modal representation, we present simulation results for the limiting case where the spot size of the modes is constant and for zero curvature. We then showed that the expected diffraction and curvature of the optical field is reproduced for vacuum propagation when a sufficient number of modes are included in the simulations using the MINERVA simulation code [2]. In addition, an analytic condition for determining the number of modes required to achieve completeness of the modal representation is given. Simulations are also presented for one short wavelength free-electron laser configuration. The MINERVA simulations are compared with simulations using the GENESIS simulation code which uses a grid-based field solver, and good agreement is found between the results obtained with both codes. We also showed that the fixed spot size/zero curvature modal representation would be useful for the simulation of the LCLS. As a consequence, we conclude that the modal representation with fixed spot size and zero curvature is an attractive option in simulating free-electron lasers.

## Acknowledgements

We would like to thank P.J.M. van der Slot for many useful and productive discussions.

**Appendix: Decomposition of a vacuum-diffracting Gaussian beam on a basis of fixed-width Gauss-Laguerre modes**

We now consider the vacuum-diffracting Gaussian optical field centered on the axis of symmetry defined earlier by

$$\delta \mathbf{A}(r,\theta,z) = \delta A_0 \frac{\hat{\mathbf{e}}_x + i\hat{\mathbf{e}}_y}{\sqrt{2}} \frac{w_0}{w} \exp\left(-\frac{r^2}{w^2}\right) \exp\left[i\left(kz + \alpha \frac{r^2}{w^2} - \psi(z)\right)\right] + c.c. ,\quad (33)$$

In this Appendix, we will derive the decomposition of this field into a basis formed by Gauss-Hermite modes with a fixed spot size and zero curvature. Defining the fixed spot size of this basis set as $\bar{w}$, we write the generic Gauss-Laguerre modal expansion as

$$\delta \mathbf{A}(r,\theta,z) = \frac{\hat{\mathbf{e}}_x + i\hat{\mathbf{e}}_y}{\sqrt{2}} \sum_{\substack{n=0 \\ l=-\infty}}^{\infty} \delta A_{nl}(z) u_{nl}(r,\theta,z) + c.c., \quad (A1)$$

where

$$u_{nl}(r,\theta,z) = \frac{1}{\bar{w}} \zeta^{|l|} L_n^{|l|}(\zeta^2) \exp\left(-\frac{r^2}{\bar{w}^2}\right) \exp\left[ikz + il\theta - i(|l| + 2n + 1)\psi(z)\right], \quad (A2)$$

and $\zeta = \sqrt{2}r/\bar{w}$. As a result,

$$\delta A_{nl}(z) = C_{nl}^2 \int_0^{2\pi} d\theta \int_0^{\infty} dr\, r\, \delta A(r,\theta,z) u_{nl}^*(r,\theta,z)$$

$$= \delta A_0 C_{nl}^2 \frac{w_0}{w\bar{w}} \exp\left[-i(|l|+2n)\psi\right] \int_0^{2\pi} d\theta \int_0^\infty dr\, r\, \zeta^{|l|} L_n^{|l|}(\zeta^2)$$

$$\times \exp\left(il\theta + i\alpha \frac{r^2}{w^2}\right) \exp\left[-r^2\left(\frac{1}{w^2} + \frac{1}{\bar{w}^2}\right)\right]$$

$$= 2\pi \delta_{l,0} C_{nl}^2 \delta A_0 \frac{w_0}{w\bar{w}} \exp(-2in\psi) \int_0^\infty dr\, r\, L_n^0(\zeta^2) \exp\left(i\alpha \frac{r^2}{w^2}\right)$$

$$\times \exp\left[-r^2\left(\frac{1}{w^2} + \frac{1}{\bar{w}^2}\right)\right], \quad \text{(A3)}$$

where $C_{nl}^2 = (2/\pi)(n!/(n+l)!)$. Observe that the only azimuthal modes that couple to the $\text{TEM}_{00}$ Gaussian mode is for $l = 0$. Changing variables $r \to \rho = \zeta^2$, we write this coefficient as

$$\delta A_{nl}(z) = 2\pi \delta_{l,0} \delta A_0 \frac{w_0 \bar{w}}{w} \exp(-2in\psi) \int_0^\infty d\rho \exp(-\chi\rho) L_n^0(\rho), \quad \text{(A4)}$$

where

$$\chi = \frac{1}{2}\left[1 + \frac{\bar{w}^2}{w^2} + i\alpha \frac{\bar{w}^2}{w^2}\right]. \quad \text{(A5)}$$

Substitution of the explicit representation for the associated Laguerre polynomials yields

$$\delta A_{nl}(z) = \delta_{l,0} C_{nl}^2 \frac{\pi \delta A_0 w_0 \bar{w}}{2w} \exp(-2in\psi) \sum_{i=0}^n (-1)^i \frac{\Gamma(n+1)}{\Gamma(n-i+1)\Gamma(i+1)} \int_0^\infty d\rho \exp(-\chi\rho) \rho^i$$

$$= \delta_{l,0} C_{nl}^2 \frac{\pi \delta A_0 w_0 \bar{w}}{2w} \exp(-2in\psi) \sum_{i=0}^n (-1)^i \frac{\Gamma(n+1)}{\Gamma(n-i+1)\Gamma(i+1)} \chi^{-i-1}$$

$$= \delta_{l,0} C_{nl}^2 \frac{\pi \delta A_0 w_0 \bar{w}}{2w} \exp(-2in\psi) \frac{1}{\chi}\left(1 - \frac{1}{\chi}\right)^n. \quad \text{(A6)}$$

It follows from this that

$$\int_0^{2\pi} d\theta \int_0^\infty dr\, r \left|\delta A_{nl} u_{nl}(r,\theta,z)\right|^2 = \delta_{l,0} \frac{\pi w_0^2 |\delta A_0|^2}{2} \frac{\bar{w}^2}{w^2} \frac{1}{\chi\chi^*}\left[\left(1 - \frac{1}{\chi}\right)\left(1 - \frac{1}{\chi^*}\right)\right]^n. \quad \text{(A7)}$$

Summing over all modes, it is possible to show that

$$\sum_{\substack{n=0 \\ l=-\infty}}^{\infty} \int_0^{2\pi} d\theta \int_0^{\infty} dr\, r \left|\delta A_{nl} u_{nl}(r,\theta,z)\right|^2 = \frac{\pi w_0^2 |\delta A_0|^2}{2} \frac{\bar{w}^2}{w^2} \frac{1}{\chi\chi^*} \left[1 - \left(1-\frac{1}{\chi}\right)\left(1-\frac{1}{\chi^*}\right)\right]^{-1}$$

$$= \frac{\pi w_0^2 |\delta A_0|^2}{2} \frac{\bar{w}^2}{w^2} \frac{1}{\chi+\chi^*-1}$$

$$= \frac{\pi w_0^2 |\delta A_0|^2}{2}. \tag{A8}$$

Comparison with Eq. (33), therefore, shows that the fixed spot size/zero curvature modal super-position reproduces the power in the TEM$_{00}$ Gaussian mode. Note that in the above derivation interference terms of the form $\int_0^{2\pi} d\theta \int_0^{\infty} dr\, r\, \delta A_{nl} \delta A_{mr}^* u_{nl} u_{mr}^*$ always integrate to zero, because of the orthogonality of the basis.

We now address the question of whether the fixed spot size/zero curvature modal expansion and dynamical equations can reproduce the development of both vacuum expansion and curvature of the TEM$_{00}$ mode. Substitution of Eq. (A3) into (A1) shows that

$$\delta \mathbf{A}(r,\theta,z) = \frac{\hat{\mathbf{e}}_x + i\hat{\mathbf{e}}_y}{\sqrt{2}} \delta A_0 \frac{w_0}{w} \exp\left(-\frac{r^2}{\bar{w}^2}\right) \exp(ikz - i\psi) \sum_{n=0}^{\infty} \frac{1}{\chi}\left(1-\frac{1}{\chi}\right)^n L_n^0(\zeta^2) + c.c.$$
(A9)

The summation over $n$ can be performed analytically by noting that

$$\sum_{n=0}^{\infty} t^n L_n^0(x) = \frac{1}{1-t} \exp\left(-\frac{tx}{1-t}\right). \tag{A10}$$

It now follows that

$$\delta \mathbf{A}(r,\theta,z) = \frac{\hat{\mathbf{e}}_x + i\hat{\mathbf{e}}_y}{\sqrt{2}} \delta A_0 \frac{w_0}{w} \exp\left(-\frac{r^2}{w^2}\right) \exp\left(ikz - i\alpha\frac{r^2}{w} - i\psi\right) + c.c.,$$

(A11)

which recovers the TEM$_{00}$ Gaussian mode. This demonstrates that the constant spot size/zero curvature modal super-position represents a complete set.